\documentclass{Interspeech}
\usepackage{multirow}
\usepackage[table]{xcolor}
\usepackage{xcolor} 
\usepackage{amsfonts} 
\usepackage{amssymb}
\usepackage{pifont}
\newcommand{\cmark}{\ding{51}}%
\newcommand{\xmark}{\ding{55}}%
\usepackage{bm}   

\interspeechcameraready 

\title{
EmoSphere-SER: Enhancing Speech Emotion Recognition Through Spherical Representation with Auxiliary Classification
}

\author[]{Deok-Hyeon}{Cho$^{*}$}
\author[]{Hyung-Seok}{Oh$^{*}$}
\author[]{Seung-Bin}{Kim$^{*}$}
\author[affiliation={{\dagger}}]{Seong-Whan}{Lee}
\affiliation[nocounter]{Department of Artificial Intelligence}{Korea University}{Seoul, Korea}

\email{dh\_cho@korea.ac.kr, hs\_oh@korea.ac.kr, sb-kim@korea.ac.kr \thanks{*Equal contribution}, sw.lee@korea.ac.kr \thanks{$^\dagger$Corresponding author}}
\keywords{Speech emotion recognition, dimensional emotion, affective computing}

\usepackage{comment}

\begin{document}

\maketitle

\begin{abstract}
Speech emotion recognition predicts a speaker's emotional state from speech signals using discrete labels or continuous dimensions such as arousal, valence, and dominance (VAD).
We propose EmoSphere-SER, a joint model that integrates spherical VAD region classification to guide VAD regression for improved emotion prediction.
In our framework, VAD values are transformed into spherical coordinates that are divided into multiple spherical regions, and an auxiliary classification task predicts which spherical region each point belongs to, guiding the regression process. 
Additionally, we incorporate a dynamic weighting scheme and a style pooling layer with multi-head self-attention to capture spectral and temporal dynamics, further boosting performance.
This combined training strategy reinforces structured learning and improves prediction consistency.
Experimental results show that our approach exceeds baseline methods, confirming the validity of the proposed framework.
\end{abstract}

\section{Introduction}
Speech emotion recognition (SER) predicts a speaker's emotional state using acoustic features \cite{ELAYADI2011572}.
Understanding the emotional state of speech is important in applications such as human-computer interaction, virtual assistants, mental health monitoring, and conversational artificial intelligence of advances in deep learning \cite{thiripurasundari2024speech, kim2015abstract, lee2018deep, lee2020continuous}.
% yun2025flowhigh, oh2025durflex

There are two main approaches to representing emotions in SER. 
The first is categorical emotion classification \cite{10444902, khan2024mser}. 
Emotion labels correspond very closely to the categories that we use in our daily lives. Paul Ekman \cite{ekman1986new} derived six primary emotions: happiness, anger, disgust, sadness, anxiety, and surprise based on universally recognized facial expressions.
This method is straightforward and easy to implement, as it reduces the complex nature of emotions to a set of clear categories.
Traditional machine learning techniques, like support vector machines and hidden Markov models, have been used to develop these systems \cite{1527805, 6023178}. 
They rely on carefully designed features extracted from speech, such as pitch, tone, and rhythm, to classify emotions effectively.
However, this approach overlooks the nuanced variations of emotions. 
The second approach is dimensional emotion prediction \cite{atmaja2020dimensional, mote24_interspeech}. 
Rather than using fixed categories, this method represents emotions along continuous scales \cite{russell1977evidence}.
Common dimensions include valence, arousal, and dominance (VAD). 
This framework offers a more detailed representation of emotional states, capturing subtle variations that categorical labels might miss.
Most emotional attribute prediction methods employ direct regression models that estimate VAD as separate numerical values \cite{10089511, Atmaja_Akagi_2020}. 
Although straightforward, this method does not fully account for the structured nature of the emotional space.
Since emotions exist in a continuous space rather than as isolated points, the lack of structured predictions in current models increases the likelihood of unnatural or conflicting outputs.

Traditional SER systems relied on a variety of extracted features to capture the nuances of speech. 
The researchers employed methods such as Mel-frequency cepstral coefficients \cite{milton2013svm}, principal component analysis \cite{CHEN20121154}, wavelet features \cite{6514336}, and Fourier parameters \cite{7009997} to describe different aspects of the speech signal, including its spectral, prosodic, and temporal characteristics.
Recent advances in SER have boosted prediction accuracy through the use of self-supervised learning models (SSL), multi-modal fusion, and deep neural networks \cite{10444902}. 
Models such as WavLM \cite{chen2022wavlm}, HuBERT \cite{9585401}, and Wav2Vec 2.0 \cite{NEURIPS2020_92d1e1eb} extract rich representations from speech, and multimodal strategies \cite{9414654} that incorporate transcript text further enhance performance. 
Although feature extraction methods have advanced the field, many models still struggle to capture the full range of emotional expressions. 
We hypothesize that incorporating a structured representation of emotion into SER models helps them learn more stable features and produce more consistent predictions.

In psychology, Reisenzein demonstrated that using the angle and length of the vector in polar coordinates is the only possible option for interpreting the relationships between emotions \cite{reisenzein1994pleasure, jenke2018cognitive}. 
Building on this idea, we introduce EmoSphere-SER—a novel approach that explicitly models emotional space using a spherical representation inspired by \cite{cho24_interspeech, 10965917} and applies spherical regions as an auxiliary loss.  
While prior work such as \cite{app15020623} has used an auxiliary classification task to predict discrete emotion categories, our approach focuses on predicting the region on the sphere, thereby emphasizing auxiliary learning for dimensional representations.
Additionally, we employ a dynamic weighting scheme to adaptively adjust the contributions of the regression and classification tasks. Furthermore, we integrate a style pooling layer to capture key spectral and temporal details of speech. These components work jointly to enhance the coherence and stability of the learned representations, paving the way for more reliable emotion recognition.
As a result, our method outperforms baseline approaches, confirming the validity of the proposed framework.

\begin{figure*}[!t] 
    \centering
\includegraphics[width=1.0\linewidth]{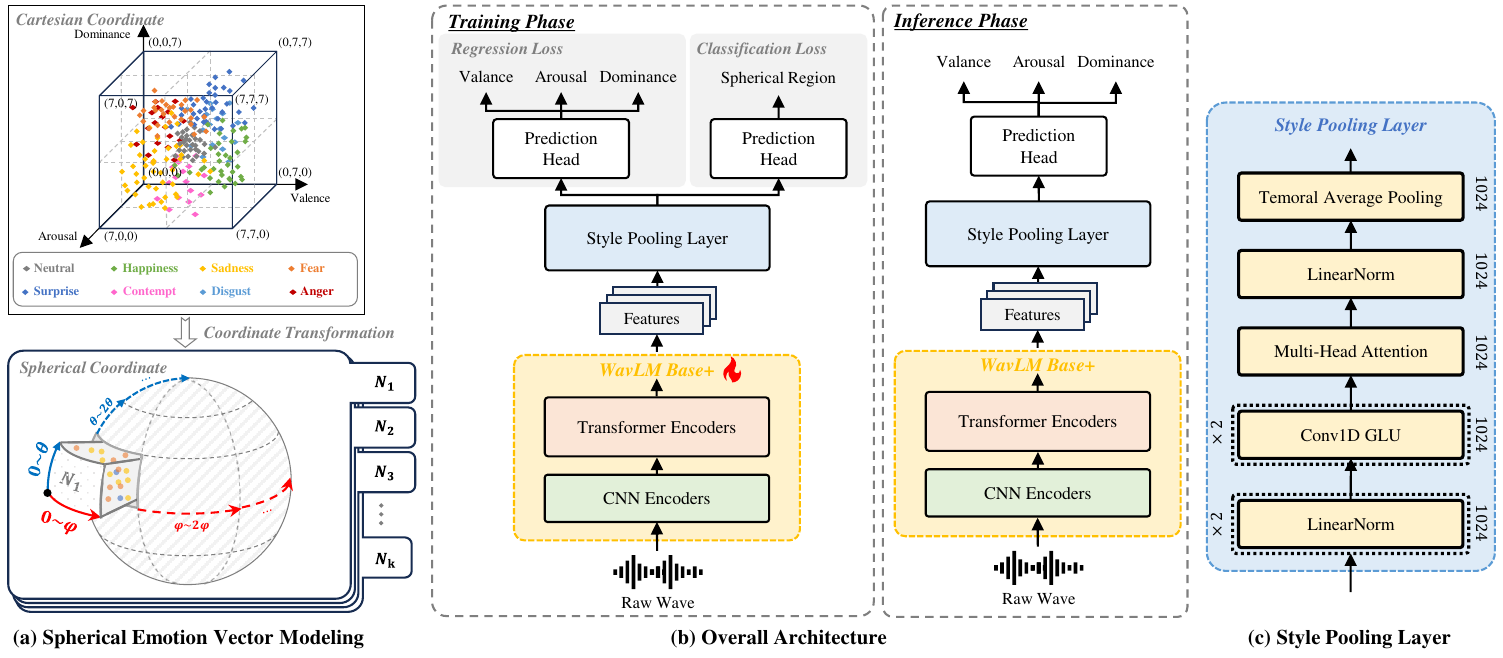}\vspace{-0.2cm}
\caption{Overall framework of EmoSphere-SER}
\label{fig1}\vspace{0cm}
\end{figure*}

\section{EmoSphere-SER}
EmoSphere-SER is an emotional attribute prediction system that employs a spherical representation to explicitly model emotional space, enabling auxiliary learning for dimensional representations. The implementation details are described in the following subsections.

\subsection{Spherical emotion vector modeling}
In this section, we present a spherical emotion vector modeling inspired by \cite{cho24_interspeech, 10965917} to enhance VAD prediction. The approach to building the space is structured around two key components: 1) normalization and spherical transformation and 2) spherical region partitioning and label assignment.

\subsubsection{Normalization and spherical transformation}
The original VAD annotations are provided on a $[1,7]$ scale. To facilitate coordinate transformation, we first normalize each of the three emotion dimensions to the range $[-1, 1]$. Once normalized, the three-dimensional VAD vector is converted from Cartesian coordinates to spherical coordinates. This transformation decomposes the vector into a magnitude of $r$, azimuth of $\phi$, and elevation of $\theta$.

\subsubsection{Spherical region partitioning and label assignment}
Inspired by the modeling of complex emotional characteristics using relative distance and angular vectors, we assume that the angle from the center determines emotional style. To implement this in the spherical coordinate system, we quantify azimuth and elevation angles to partition the space into distinct spherical regions. The azimuth angle is divided into $N_{\phi}$ equal intervals, while the elevation angle is divided into $N_{\theta}$ equal intervals, resulting in $N = N_{\phi}  \times  N_{\theta}$ regions. Each region is then assigned a unique categorical label corresponding to a prototypical emotional state within the affective space. 
% To leverage this discrete structure, we introduce an auxiliary loss function that encourages the learned vector representations to be aligned with the appropriate spherical sector. The network is explicitly regularized so that its latent representation is positioned in the sector that best reflects its emotional characteristics. This alignment is expected to facilitate a more robust and structured VAD prediction. 

\subsection{Architecture}
In this section, we introduce the overall architecture of emotional attribute prediction based on a pre-trained SSL model. The framework consists of three main components: a pre-trained SSL model, a style pooling layer, and prediction heads, which will be explained in detail in the following subsections.

\subsubsection{Pre-trained SSL model}
We adopt WavLM \cite{chen2022wavlm} as our feature encoder for extracting features from audio. WavLM is a pre-trained SSL model trained on large-scale data, and it is capable of extracting frame-level high-dimensional representations. To further enhance the model's ability to capture emotional nuances in speech, we fine-tuned WavLM specifically for emotion recognition.

\subsubsection{Style pooling layer}
To aggregate frame-level features into an utterance-level representation, we use a style pooling layer inspired by \cite{pmlr-v139-min21b}. This layer consists of LayerNorm, Conv1D with GLU, and multi-head attention, allowing the model to attend to different temporal segments simultaneously. Finally, we apply temporal average pooling to summarize the frame-level features.

\subsubsection{Prediction head}
From the utterance-level representations generated by style pooling layer, we employ two distinct prediction layers to forecast both the spherical region and the continuous VAD values. To predict the spherical region, a classification layer maps the learned representations to one of the ${N}$ discrete categories. Simultaneously, a regression layer is implemented to output three scalar values, each corresponding to one of the VAD dimensions.
% To predict the spherical region, we utilize a classification layer that maps the learned representations to one of ${N}$ discrete categories, each corresponding to a specific sector derived from the spherical transformation of the normalized VAD values. In parallel, a dedicated regression layer is implemented to predict the continuous values of VAD. This layer outputs three scalar values, each corresponding to one of the VAD dimensions. By directly regressing the VAD values, the model produces the final emotion predictions.

\begin{table*}[!ht]
    \centering 
        \caption{Performance comparison of our proposed method under different experimental settings. A {\ding{55}} symbol indicates that the corresponding component was excluded from the model. }
    \label{Table1}\vspace{-0.2cm}
        \resizebox{1.0\textwidth}{!}{
    \begin{tabular}{l|>{\centering\arraybackslash}p{3.0cm}>{\centering\arraybackslash}p{3.0cm}>{\centering\arraybackslash}p{3.0cm}|ccc|c}
        \toprule
        \textbf{Method} & \textbf{Auxiliary Loss} & \textbf{Style Pooling Layer} & \textbf{Data Preprocessing} & \textbf{Valance} & \textbf{Arousal} & \textbf{Dominance} & \textbf{Average} \\ 
        \midrule
            EmoSphere-SER & \cmark & \cmark & \cmark & 0.6952 & \textbf{0.7482} & \textbf{0.6220} & \textbf{0.6884} \\ 
        \midrule
            \multirow{3}{10em}{Ablation Study} & \cmark & \cmark & \xmark $\cellcolor{gray!25}$ & \textbf{0.6953} & 0.7447 & 0.6166 & 0.6855 \\ 
            & \cmark & \xmark $\cellcolor{gray!25}$ & \xmark $\cellcolor{gray!25}$ & 0.6908 & 0.7456 & 0.6192 & 0.6852 \\ 
            & \xmark $\cellcolor{gray!25}$ & \xmark $\cellcolor{gray!25}$ & \xmark $\cellcolor{gray!25}$ & 0.6845 & 0.7349 & 0.6210 & 0.6801 \\ 
        \bottomrule
    \end{tabular}
      }\vspace{0.0cm}
\end{table*}

\subsection{Training strategy}
Our training strategy leverages an auxiliary spherical classification loss and the primary VAD regression loss.  The spherical region classification loss is designed to support VAD prediction.

\subsubsection{Spherical region classification loss}
To facilitate the alignment of the utterance-level representations with the ${N}$ discrete spherical regions, we empoly an auxiliary classification loss, denoted as $\mathcal{L}_{\text{sph}}$. This loss is computed via a weighted cross-entropy (WCE) loss between the predicted and ground-truth spherical region labels, as the following equation:
\begin{equation}
    \mathcal{L}_{\text{sph}}= - \sum_{i=1}^{{N}}w_{i} y_{i} \text{log}\left ( p_{i} \right ),
\end{equation}
where $i$ represents index of each class and $p_{i}$ is the predicted probability that the sample belongs to class $i$. The variable $y_{i}$ is a one-hot vector representing the true class. The $w_{i}$ is the weight assigned to class $i$, determined based on its inverse frequency to mitigate class imbalance.

\subsubsection{VAD regression loss}
For the primary task of VAD prediction, we adopt the concordance correlation coefficient (CCC) loss, which directly optimizes the agreement between the continuous predicted values and the ground-truth annotations. Given the predicted VAD values $\hat{y}$ and the corresponding ground-truth $y$, the $\mathcal{L}_{CCC}$ is defined as:
\begin{equation}
    \mathcal{L}_{CCC} = \frac{2\sigma_{y\hat{y}}}{\sigma_y^2 + \sigma_{\hat{y}}^2 + (\mu_y - \mu_{\hat{y}})^2},
\end{equation}
where $\mu_y$ and $\mu_{\hat{y}}$ denote the means of $y$ and $\hat{y}$, respectively, $\sigma_y^2$ and $\sigma_{\hat{y}}^2$ are the variances, $\sigma_{y\hat{y}}$ represents the covariance between $y$ and $\hat{y}$.

\subsubsection{Overall loss}
The complete training objective combines both the auxiliary spherical classification loss and the VAD regression loss:
\begin{equation}
    \mathcal{L} = \mathcal{L}_{CCC} + \lambda_{sph}\mathcal{L}_{sph},
\end{equation}
where $\lambda_{sph}$ denotes a weighting coefficient that controls the contribution of the auxiliary loss. During the initial training phase, $\lambda_{sph}$ decays linearly, and after a certain point, the model is trained solely with $\mathcal{L}_{CCC}$. Specifically, the coefficient $\lambda_{sph}$ is defined as follows:
\begin{equation}
    \lambda_{sph} =
\begin{cases}
1 - \frac{0.99}{5}e, & \text{if } 0 \leq e < 5, \\
0, & \text{if } e \geq 5,
\end{cases}
\end{equation}
where $e$ denotes the number of epoch.

While the SSL model and style pooling layer are shared between both learning tasks, the remaining components are fundamentally separate. Notably, the spherical region classification loss converges rapidly, allowing the model to shift its focus to VAD regression once a certain level of convergence is achieved. This enables more fine-grained predictions of emotional attributes, with empirical results showing that performance peaks at around 5 epochs.

\section{Experiments and results}
\subsection{Experimental setup}
We utilized the MSP-Podcast corpus dataset \cite{lotfian2017building}, which comprises approximately 237 hours of speech data annotated with both categorical emotion labels and dimensional VAD values. Each utterance has been perceptually annotated by at least five raters. The dataset includes eight categorical emotion classes: happiness, sadness, fear, surprise, contempt, disgust, and a neutral state. For dimensional emotion labels, raters evaluated arousal, dominance, and valence using a seven-point Likert scale. 
% The dataset is divided into three subsets: the training set consists of 68,119 samples, the development set includes 19,815 speaking segments from 454 speakers, and the test set contains 2,347 unique segments from 187 speakers. The labels for the test set have not been publicly released as part of the challenge. 
We adhered to the challenge organizers' guidelines for data partitioning, utilizing the recommended training split for model development \cite{goncalves2024odyssey}. Additionally, to ensure a balanced representation of categories within the validation set, we further divided the original validation set into a validation set and an in-house test set. Specifically, the validation set was constructed by sampling 300 instances per category, maintaining uniform class distribution, while the remaining data was designated as the in-house test set. The ``X'' label corresponds to instances where no plurality voting winner could be determined among the emotion categories, reflecting cases of annotator disagreement or ambiguity. To minimize label noise and ensure more reliable supervision, all instances with the ``X'' label were excluded from the training, validation, and in-house test sets.

We utilize the AdamW optimizer \cite{loshchilov2017decoupled} with a learning rate of $1\times 10^{-5}$ for training. The model is trained for 20 epochs with a batch size of 32. The model checkpoint corresponding to the best validation performance on the development set is selected and saved based on the lowest validation loss. The training process of the TTS module was conducted over approximately 24 hours on a single NVIDIA RTX A6000 GPU.

\subsection{Implementation details}
The style pooling layer is based on the structure of the style encoder proposed in \cite{min2021meta} to embed the reference speech into a latent vector. It consists of three main components: The spectral processing module includes two fully connected layers with 1024 hidden units each. The temporal processing module is composed of two gated 1D convolutional neural networks with residual connections, where the convolutional layers have a filter size of 1024 and a kernel size of 5. Following this, the multi-head self-attention module has a hidden size of 1024 with two attention heads. On top of this module, a fully connected layer with a dimensionality of 1024 is applied, followed by temporal average pooling.
% The processed output from the style pooling layer is subsequently fed into a series of FC layers for the final prediction. 

\subsection{Performance Metrics}
To evaluate the model’s performance across different emotion representations, we use metrics designed for each type. For spherical emotion vector representations, F1-score and accuracy are used to evaluate model performance, offering a comprehensive view of both per-class balance and overall predictive correctness. For dimensional emotion representations, the mean concordance correlation coefficient (CCC) assesses how well the model predicts continuous emotional states. This combination of metrics provides a comprehensive evaluation of the model’s ability to capture both discrete spherical regions and the continuous nature of emotions.

\subsection{Model performance}
As shown in Table \ref{Table1}, our model improves performance, explained as follows: 1) ``\textbf{w/o Data Preprocessing}'' indicates that the model is trained on the entire dataset without excluding category data with the ``X'' label. By removing instances with multiple speakers or ambiguous emotions, the model achieved better performance in predicting emotional attributes. 2) ``\textbf{w/o Style Pooling Layer}'' denotes to attentive statistics pooling \cite{okabe18_interspeech}, which employs an attention mechanism to assign different weights to frames. By replacing attentive statistics pooling with the style pooling layer, our model benefits from a more expressive style representation. Unlike traditional pooling methods, the style pooling layer effectively captures both spectral and temporal dynamics while leveraging multi-head self-attention to enhance feature extraction. This allows the model to generate richer and more speaker-adaptive embeddings, ultimately leading to improved performance in predicting emotional attributes. 3) ``\textbf{w/o Auxiliary Loss}'' indicates the absence of an additional auxiliary loss designed to enhance VAD prediction. Specifically, we introduce an auxiliary loss that converts the original VAD values into a spherical emotion vector representation, predicting its corresponding region in spherical coordinates using WCE as a categorical loss. This auxiliary objective aids the model in the early stages of training by providing additional supervision, ultimately leading to improved VAD prediction performance. These results confirm that the proposed method effectively enhances the prediction of emotional attributes and demonstrates the ability of the SER model to adapt to various emotions and speaker styles in the prediction of VAD.

\begin{table}[!t]
    \centering 
        \caption{Comparison results on auxiliary loss modifications.}
    \label{Table2}\vspace{-0.2cm}
        \resizebox{1.0\columnwidth}{!}{
    \begin{tabular}{l|ccc|c}
        \toprule
        {\textbf{Methods}} & \textbf{Valance} & \textbf{Arousal} & \textbf{Dominance} & \textbf{Average} \\ 
        \midrule
        % 원본 Dynamic Weighting scheme
         w/o Dynamic Weighting & 0.6951 & 0.7405 & 0.6267 & 0.6875 \\
         % 원본: Categorical Emotion Recognition
         w/ Categorical Recognition & 0.6912 & 0.7342 & 0.6281 & 0.6845 \\
         w/ Cross Entropy & \textbf{0.6968} & 0.7352 & \textbf{0.6308} & 0.6876 \\
        \midrule
         EmoSphere-SER (Proposed) & 0.6952 & \textbf{0.7482} & 0.6220 & \textbf{0.6884} \\
        \bottomrule
    \end{tabular}
      } % \vspace{0cm}
\end{table}

\subsection{Impact of auxiliary loss modification}
To evaluate and compare the effectiveness of our proposed auxiliary loss, we conducted three additional experiments: (1) ``\textbf{w/o Dynamic Weighting}'' refers to training without dynamically adjusting the auxiliary loss weight, instead maintaining a constant weight throughout the entire training process. (2) ``\textbf{w/ Categorical Recognition}'' replaces the spherical region classification with a categorical emotion prediction task, where the model uses categorical emotion labels as the auxiliary loss instead. (3) ``\textbf{w/ Cross Entropy}'' replaces WCE with standard cross-entropy in the spherical region classification task.

As shown in Table \ref{Table2}, the results indicate that our proposed auxiliary loss, which models VAD as a spherical region prediction task, provides more effective supervision than direct categorical emotion prediction. This suggests that learning a latent structured representation in the spherical region better aligns with VAD estimation. Additionally, the dynamic weighting scheme contributes to refining VAD predictions. By emphasizing auxiliary supervision in the early training stages and gradually shifting the focus toward VAD regression, the model achieves more precise and detailed predictions in later stages. Moreover, the use of WCE outperforms standard cross-entropy by better handling class imbalances, ensuring that each spherical region is adequately represented during training. In summary, the proposed module enhances VAD prediction by leveraging structured auxiliary supervision and dynamic weighting, leading to more precise and stable predictions. Combining both models provides a more comprehensive understanding of emotions, capturing both fundamental patterns and subtle nuances of emotional experiences.

\subsection{Impact of angular division}
As shown in Table \ref{Table3}, we investigated the impact of different angular divisions for the spherical region. We conducted experiments by dividing the total number of regions, represented as ${N}=N_{\phi}N_{\theta}$, based on angular thresholds of $90\mbox{°}$, $60\mbox{°}$, and $45\mbox{°}$. The experimental results indicate that the model achieved the best performance in both the prediction of emotional attributes and the classification of the spherical region when the angular division was set to $90\mbox{°}$  (i.e., $N_\phi = 2$, $N_\theta = 4$, and consequently $N = 8$). For angular divisions smaller than $90\mbox{°}$, the increased difficulty in predicting the spherical region made auxiliary learning more challenging, leading to a decline in performance. These findings suggest that $90\mbox{°}$ provides an optimal balance between categorical supervision and effective auxiliary learning, leading to enhanced VAD prediction.

\begin{table}[!t]
    \centering 
        \caption{Performance comparison of different angular divisions in the auxiliary loss across varying azimuth angles with identical elevation angles to ensure uniform spherical partitioning.  The total number of regions is represented as ${N}=N_{\phi}N_{\theta}$, where $N_{\phi}$ and $N_{\theta}$ denote the divisions along the azimuth and elevation axes, respectively.}
    \label{Table3}\vspace{-0.2cm}
        \resizebox{1.00\columnwidth}{!}{
    \begin{tabular}{l|ccc|c|cc}
        \toprule
        \multirow{2}{*}{$\bm{N}$ \textbf{(Angle)}} & \multicolumn{4}{c|}{\textbf{Emotional Attribute}} & \multicolumn{2}{c}{\textbf{Spherical Region}}  \\ 
        \cmidrule{2-7}
         & \textbf{Valance} & \textbf{Arousal}  & \textbf{Dominance}  & \textbf{Average}  & \textbf{Macro F1}  & \textbf{Accuracy}  \\ 
        \midrule
         32 (45°) & 0.6833 & 0.7385 & 0.6204 & 0.6808 & 4.77 & 16.95 \\
         18 (60°) & \textbf{0.6962} & 0.7376 & \textbf{0.6250} & 0.6863 & 17.96 & 31.95 \\
        \midrule
         8 (90°) & 0.6952 & \textbf{0.7482} & 0.6220 & \textbf{0.6884} & 36.27 & 59.51 \\
        \bottomrule
    \end{tabular}
      }\vspace{0cm}
\end{table}

\section{Conclusion}
In this work, we present EmoSphere-SER, a speech emotion recognition model that improves the prediction of continuous emotional attributes by incorporating a structured spherical representation of emotion. 
Our model divides the emotional space into distinct regions and employs an auxiliary classification task to identify the region corresponding to each emotional state, thereby guiding the regression process.
Furthermore, the dynamic weighting scheme maintains balance in the overall learning process, and a style pooling layer with multi-head self-attention effectively captures both spectral and temporal dynamics. 
Experimental results demonstrate that our approach outperforms baseline methods, confirming the benefits of our structured framework for enhanced emotional attribute prediction. To the best of our knowledge, this work represents the first attempt to predict emotional attributes using a spherical coordinate-based approach. Future research can further explore refining region partitioning strategies and extending this framework to other affective computing tasks.

\section{Acknowledgements}
This work was partly supported by the Institute of Information \& Communications Technology Planning \& Evaluation (IITP) grant funded by the Korea government (MSIT) (Artificial Intelligence Graduate School Program (Korea University) (No. RS-2019-II190079), Artificial Intelligence Innovation Hub (No. RS-2021-II212068), AI Technology for Interactive Communication of Language Impaired Individuals (No. RS-2024-00336673), and Artificial Intelligence Star Fellowship Support Program to Nurture the Best Talents (IITP-2025-RS-2025-02304828)).

\bibliographystyle{IEEEtran}
\bibliography{refs}

\begin{thebibliography}{10}
\providecommand{\url}[1]{#1}
\csname url@samestyle\endcsname
\providecommand{\newblock}{\relax}
\providecommand{\bibinfo}[2]{#2}
\providecommand{\BIBentrySTDinterwordspacing}{\spaceskip=0pt\relax}
\providecommand{\BIBentryALTinterwordstretchfactor}{4}
\providecommand{\BIBentryALTinterwordspacing}{\spaceskip=\fontdimen2\font plus
\BIBentryALTinterwordstretchfactor\fontdimen3\font minus \fontdimen4\font\relax}
\providecommand{\BIBforeignlanguage}[2]{{%
\expandafter\ifx\csname l@#1\endcsname\relax
\typeout{** WARNING: IEEEtran.bst: No hyphenation pattern has been}%
\typeout{** loaded for the language `#1'. Using the pattern for}%
\typeout{** the default language instead.}%
\else
\language=\csname l@#1\endcsname
\fi
#2}}
\providecommand{\BIBdecl}{\relax}
\BIBdecl

\bibitem{ELAYADI2011572}
M.~{El Ayadi}, M.~S. Kamel, and F.~Karray, ``Survey on speech emotion recognition: Features, classification schemes, and databases,'' \emph{Pattern Recognit.}, vol.~44, no.~3, pp. 572--587, 2011.

\bibitem{thiripurasundari2024speech}
D.~Thiripurasundari, K.~Bhangale, V.~Aashritha, S.~Mondreti, and M.~Kothandaraman, ``Speech emotion recognition for human--computer interaction,'' \emph{Int. J. Speech Technol.}, vol.~27, no.~3, pp. 817--830, 2024.

\bibitem{kim2015abstract}
J.~Kim, J.~Schultz, T.~Rohe, C.~Wallraven, S.-W. Lee, and H.~H. B{\"u}lthoff, ``Abstract representations of associated emotions in the human brain,'' \emph{Journal of Neuroscience}, vol.~35, no.~14, pp. 5655--5663, 2015.

\bibitem{lee2018deep}
K.~Lee, S.-A. Kim, J.~Choi, and S.-W. Lee, ``Deep reinforcement learning in continuous action spaces: a case study in the game of simulated curling,'' in \emph{Int. Conf. Mach. Learn.}, 2018.

\bibitem{lee2020continuous}
D.-H. Lee, J.-H. Jeong, K.~Kim, B.-W. Yu, and S.-W. Lee, ``Continuous eeg decoding of pilots’ mental states using multiple feature block-based convolutional neural network,'' \emph{IEEE access}, vol.~8, pp. 121\,929--121\,941, 2020.

\bibitem{10444902}
W.~Chen, X.~Xing, P.~Chen, and X.~Xu, ``Vesper: A compact and effective pretrained model for speech emotion recognition,'' \emph{IEEE Trans. Affect. Comput.}, vol.~15, no.~3, pp. 1711--1724, 2024.

\bibitem{khan2024mser}
M.~Khan, W.~Gueaieb, A.~El~Saddik, and S.~Kwon, ``Mser: Multimodal speech emotion recognition using cross-attention with deep fusion,'' \emph{Expert Syst. Appl.}, vol. 245, p. 122946, 2024.

\bibitem{ekman1986new}
P.~Ekman and W.~V. Friesen, ``A new pan-cultural facial expression of emotion,'' \emph{Motiv. Emot.}, vol.~10, pp. 159--168, 1986.

\bibitem{1527805}
Y.-L. Lin and G.~Wei, ``Speech emotion recognition based on hmm and svm,'' in \emph{Int. Conf. Mach. Learn. Cybern.}, vol.~8, 2005, pp. 4898--4901.

\bibitem{6023178}
P.~Shen, Z.~Changjun, and X.~Chen, ``Automatic speech emotion recognition using support vector machine,'' in \emph{2011 Int. Conf. Electron. Mech. Eng. Inf. Technol.}, vol.~2, 2011, pp. 621--625.

\bibitem{atmaja2020dimensional}
B.~T. Atmaja and M.~Akagi, ``Dimensional speech emotion recognition from speech features and word embeddings by using multitask learning,'' \emph{APSIPA Trans. Signal Inf. Process.}, vol.~9, p. e17, 2020.

\bibitem{mote24_interspeech}
P.~Mote, B.~Sisman, and C.~Busso, ``Unsupervised domain adaptation for speech emotion recognition using k-nearest neighbors voice conversion,'' in \emph{Interspeech}, 2024, pp. 1045--1049.

\bibitem{russell1977evidence}
J.~A. Russell and A.~Mehrabian, ``Evidence for a three-factor theory of emotions,'' \emph{J. Res. Pers.}, vol.~11, no.~3, pp. 273--294, 1977.

\bibitem{10089511}
J.~Wagner, A.~Triantafyllopoulos, H.~Wierstorf, M.~Schmitt, F.~Burkhardt, F.~Eyben, and B.~W. Schuller, ``Dawn of the transformer era in speech emotion recognition: Closing the valence gap,'' \emph{IEEE Trans. Pattern Anal. Mach. Intell.}, vol.~45, no.~9, pp. 10\,745--10\,759, 2023.

\bibitem{Atmaja_Akagi_2020}
B.~T. Atmaja and M.~Akagi, ``Dimensional speech emotion recognition from speech features and word embeddings by using multitask learning,'' \emph{APSIPA Trans. Signal Inf. Process.}, vol.~9, p. e17, 2020.

\bibitem{milton2013svm}
A.~Milton, S.~S. Roy, and S.~T. Selvi, ``Svm scheme for speech emotion recognition using mfcc feature,'' \emph{Int. J. Comput. Appl.}, vol.~69, no.~9, 2013.

\bibitem{CHEN20121154}
L.~Chen, X.~Mao, Y.~Xue, and L.~L. Cheng, ``Speech emotion recognition: Features and classification models,'' \emph{Digit. Signal Process.}, vol.~22, no.~6, pp. 1154--1160, 2012.

\bibitem{6514336}
K.~Krishna~Kishore and P.~Krishna~Satish, ``Emotion recognition in speech using mfcc and wavelet features,'' in \emph{IEEE Int. Adv. Comput. Conf. (IACC)}, 2013, pp. 842--847.

\bibitem{7009997}
K.~Wang, N.~An, B.~N. Li, Y.~Zhang, and L.~Li, ``Speech emotion recognition using fourier parameters,'' \emph{IEEE Trans. Affect. Comput.}, vol.~6, no.~1, pp. 69--75, 2015.

\bibitem{chen2022wavlm}
S.~Chen, C.~Wang, Z.~Chen, Y.~Wu, S.~Liu, Z.~Chen, J.~Li, N.~Kanda, T.~Yoshioka, X.~Xiao \emph{et~al.}, ``Wavlm: Large-scale self-supervised pre-training for full stack speech processing,'' \emph{IEEE J. Sel. Top. Signal Process.}, vol.~16, no.~6, pp. 1505--1518, 2022.

\bibitem{9585401}
W.-N. Hsu, B.~Bolte, Y.-H.~H. Tsai, K.~Lakhotia, R.~Salakhutdinov, and A.~Mohamed, ``Hubert: Self-supervised speech representation learning by masked prediction of hidden units,'' \emph{IEEE/ACM Trans. Audio, Speech, Lang. Process.}, vol.~29, pp. 3451--3460, 2021.

\bibitem{NEURIPS2020_92d1e1eb}
A.~Baevski, Y.~Zhou, A.~Mohamed, and M.~Auli, ``wav2vec 2.0: A framework for self-supervised learning of speech representations,'' in \emph{Adv. Neural Inf. Process. Syst.}, vol.~33, 2020, pp. 12\,449--12\,460.

\bibitem{9414654}
L.~Sun, B.~Liu, J.~Tao, and Z.~Lian, ``Multimodal cross- and self-attention network for speech emotion recognition,'' in \emph{IEEE Int. Conf. Acoust., Speech, Signal Process. (ICASSP)}, 2021, pp. 4275--4279.

\bibitem{reisenzein1994pleasure}
R.~Reisenzein, ``Pleasure-arousal theory and the intensity of emotions,'' \emph{J. Pers. Soc. Psychol.}, vol.~67, no.~3, p. 525, 1994.

\bibitem{jenke2018cognitive}
R.~Jenke and A.~Peer, ``A cognitive architecture for modeling emotion dynamics: Intensity estimation from physiological signals,'' \emph{Cogn. Syst. Res.}, vol.~49, pp. 128--141, 2018.

\bibitem{cho24_interspeech}
D.-H. Cho, H.-S. Oh, S.-B. Kim, S.-H. Lee, and S.-W. Lee, ``Emosphere-tts: Emotional style and intensity modeling via spherical emotion vector for controllable emotional text-to-speech,'' in \emph{Interspeech}, 2024, pp. 1810--1814.

\bibitem{10965917}
D.-H. Cho, H.-S. Oh, S.-B. Kim, and S.-W. Lee, ``Emosphere++: Emotion-controllable zero-shot text-to-speech via emotion-adaptive spherical vector,'' \emph{IEEE Trans. Affect. Comput.}, pp. 1--16, 2025.

\bibitem{app15020623}
J.~L. Bautista and H.~S. Shin, ``Speech emotion recognition model based on joint modeling of discrete and dimensional emotion representation,'' \emph{Appl. Sci.}, vol.~15, no.~2, 2025.

\bibitem{pmlr-v139-min21b}
D.~Min, D.~B. Lee, E.~Yang, and S.~J. Hwang, ``Meta-stylespeech : Multi-speaker adaptive text-to-speech generation,'' in \emph{Int. Conf. Mach. Learn.}, vol. 139, 2021, pp. 7748--7759.

\bibitem{lotfian2017building}
R.~Lotfian and C.~Busso, ``Building naturalistic emotionally balanced speech corpus by retrieving emotional speech from existing podcast recordings,'' \emph{IEEE Trans. Affect. Comput.}, vol.~10, pp. 471--483, 2017.

\bibitem{goncalves2024odyssey}
L.~Goncalves, A.~N. Salman, A.~R. Naini, L.~M. Velazquez, T.~Thebaud, L.~P. Garcia, N.~Dehak, B.~Sisman, and C.~Busso, ``Odyssey 2024-speech emotion recognition challenge: Dataset, baseline framework, and results,'' \emph{Development}, vol.~10, no. 9,290, pp. 4--54, 2024.

\bibitem{loshchilov2017decoupled}
I.~Loshchilov and F.~Hutter, ``Decoupled {W}eight {D}ecay {R}egularization,'' in \emph{Int. Conf. Learn. Represent.}, 2019.

\bibitem{min2021meta}
D.~Min, D.~B. Lee, E.~Yang, and S.~J. Hwang, ``Meta-stylespeech: Multi-speaker adaptive text-to-speech generation,'' in \emph{Int. Conf. Mach. Learn.}, 2021, pp. 7748--7759.

\bibitem{okabe18_interspeech}
K.~Okabe, T.~Koshinaka, and K.~Shinoda, ``Attentive statistics pooling for deep speaker embedding,'' in \emph{Interspeech}, 2018, pp. 2252--2256.

\end{thebibliography}

\end{document}